\documentclass[superscriptaddress,amsmath,amssymb,aps,prl,twocolumn]{revtex4-1}

\usepackage{times}
\usepackage{graphicx}
\usepackage{dcolumn}

\begin{document}

\title{Angle-multiplexed metasurfaces: encoding independent wavefronts in a single metasurface under different illumination angles}

\author{Seyedeh Mahsa Kamali}
\affiliation{T. J. Watson Laboratory of Applied Physics and Kavli Nanoscience Institute, California Institute of Technology, 1200 E California Blvd., Pasadena, CA 91125, USA}
\author{Ehsan Arbabi}
\affiliation{T. J. Watson Laboratory of Applied Physics and Kavli Nanoscience Institute, California Institute of Technology, 1200 E California Blvd., Pasadena, CA 91125, USA}
\author{Amir Arbabi}
\affiliation{Department of Electrical and Computer Engineering, University of Massachusetts Amherst, 151 Holdsworth Way, Amherst, MA 01003, USA}
\author{Yu Horie}
\affiliation{T. J. Watson Laboratory of Applied Physics and Kavli Nanoscience Institute, California Institute of Technology, 1200 E California Blvd., Pasadena, CA 91125, USA}
\author{MohammadSadegh Faraji-Dana}
\affiliation{T. J. Watson Laboratory of Applied Physics and Kavli Nanoscience Institute, California Institute of Technology, 1200 E California Blvd., Pasadena, CA 91125, USA}
\author{Andrei Faraon}
\email{Corresponding author: A.F: faraon@caltech.edu}
\affiliation{T. J. Watson Laboratory of Applied Physics and Kavli Nanoscience Institute, California Institute of Technology, 1200 E California Blvd., Pasadena, CA 91125, USA}

\begin{abstract}
The angular response of thin diffractive optical elements is highly correlated. For example, the angles of incidence and diffraction of a grating are locked through the grating momentum determined by the grating period. Other diffractive devices, including conventional metasurfaces, have a similar angular behavior due to the fixed locations of the Fresnel zone boundaries and the weak angular sensitivity of the meta-atoms. To alter this fundamental property, we introduce {\it angle-multiplexed metasurfaces}, composed of reflective high-contrast dielectric U-shaped meta-atoms, whose response under illumination from different angles can be controlled independently. This enables flat optical devices that impose different and independent optical transformations when illuminated from different directions, a capability not previously available in diffractive optics.

\end{abstract}

\maketitle

\begin{figure*}[htp]
\centering
\includegraphics[width=2\columnwidth]{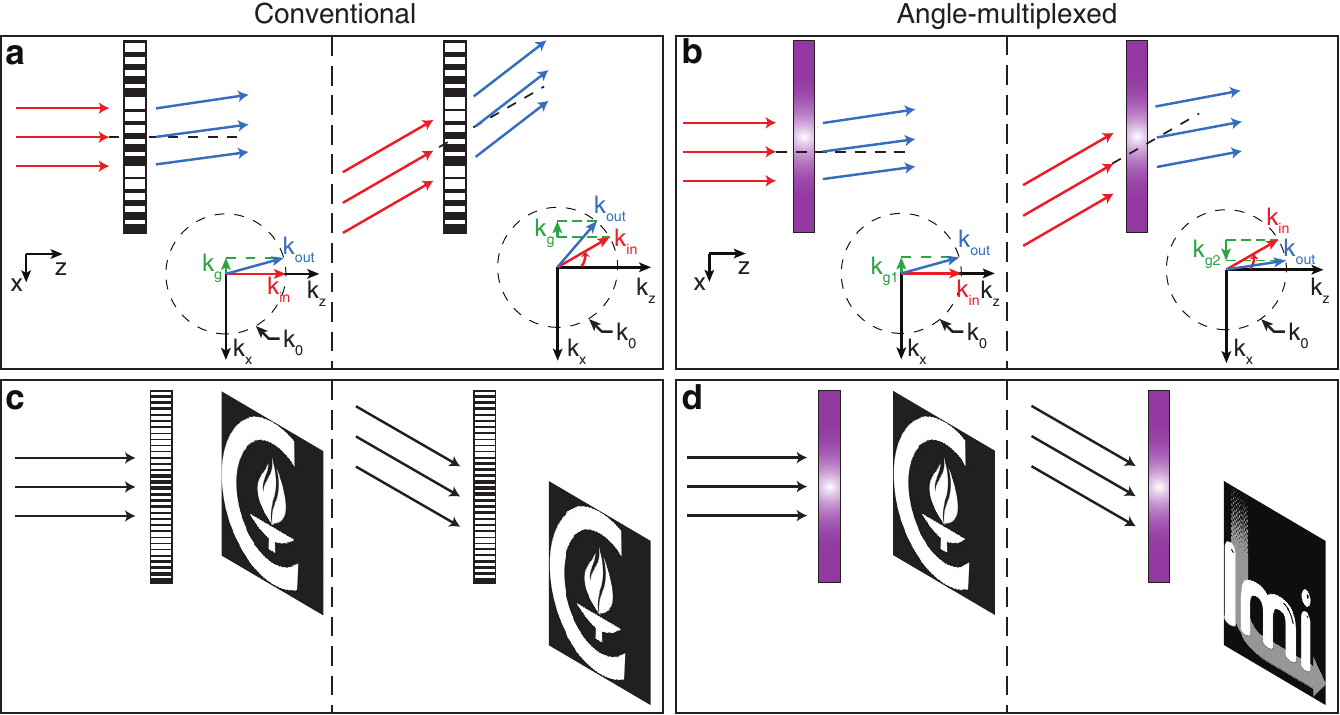}
\caption{\textbf{Angle-multiplexed metasurface concept}. \textbf{a}, Schematic illustration of diffraction of light by a grating. A grating adds a fixed linear momentum ($\hbar k_g$) to the incident light, independent of the illumination angle. If the illumination angle deviates from the designated incident angle, light is deflected to a different angle which is dictated by the grating period. \textbf{b}, Illustration of the angle-multiplexed metasurface platform. This platform provides different responses according to the illumination angle. For instance, two gratings with different deflection angles (different grating momenta) can be multiplexed such that different illumination angles acquire different momenta. \textbf{c}, Illustration of a typical hologram that creates one specific image (Caltech logo) under one illumination angle (left). The same hologram will be translated laterally (and distorted) by tilting the illumination angle (right). \textbf{d}, Schematic illustration of an angle-multiplexed hologram. Different images are created under different illumination angles. For ease of illustration, the devices are shown in transmission while the actual fabricated devices are designed to operate in reflection mode.}
\label{fig:1_Concept}
\end{figure*}

The concept of angular correlation is schematically depicted in Fig. \ref{fig:1_Concept}a for a diffraction grating. In gratings, the diffraction angle $\theta_m$ of order $m$ is related to the incident angle $\theta_\mathrm{in}$ by the relation $d(\sin(\theta_m)-\sin(\theta_\mathrm{in}))=m\lambda$, where $\lambda$ is the wavelength, and $d$ is the grating period, determined solely by the geometry. Therefore, a grating adds a fixed ``linear momentum", dictated by its period, to the momentum of the incident light regardless of the incident angle. Similarly, a regular hologram designed to project a certain image when illuminated from a given angle will project the same image (with possible distortions and efficiency reduction) when illuminated from a different angle (Fig. \ref{fig:1_Concept}c). The concept that we introduce here is shown schematically in Fig. \ref{fig:1_Concept}b for an angle-multiplexed grating that adds a different ``linear momentum" depending on the angle of incidence, and Fig. \ref{fig:1_Concept}d for an angle-multiplexed hologram that displays a different image depending on the angle of incidence. Breaking this fundamental correlation and achieving independent control over distinct incident angles is conceptually new and results in the realization of a new category of compact multifunctional devices which allow for embedding several functions into a thin single metasurface.

Optical metasurfaces are two-dimensional arrangements of a large number of discrete meta-atoms that enable precise control of optical wavefronts with subwavelength resolution\cite{kildishev2013planar,Vo2014IEEEPhotonTechLett,lalanne1998blazed,Astilean1998OptLett,qiao2017recent,jahani2016all,
staude2017metamaterial,hsiao2017fundamentals,genevet2017recent,kuznetsov2016optically,
zhu2017traditional,ding2017gradient,jiang2017multifunctional}. Several devices with the ability to control the phase \cite{Arbabi2015NatCommun,yu2015high,chen2017reconfigurable,jha2015metasurface}, polarization \cite{Arbabi2015NatNano,pfeiffer2013cascaded,arbabi2016high}, and amplitude \cite{cencillo2017electro,Zhao2011Metamat,thyagarajan2017millivolt} of light have been demonstrated. They can directly replace traditional bulk optical components like gratings \cite{lin2017optical,sell2017large}, lenses \cite{Arbabi2015NatCommun,khorasaninejad2016metalenses,chen2017immersion,arbabi2016multiwavelength}, waveplates \cite{backlund2016removing,ding2015broadband,jiang2014broadband}, polarizers \cite{Arbabi2015NatNano,desiatov2015polarization}, holograms \cite{wang2016grayscale,choudhury2017pancharatnam}, orbital angular momentum generators \cite{ren2016orbital,bouchard2014optical}, or provide novel functionalities \cite{Arbabi2015NatNano,kamali2016,arbabi2017controlling,arbabi2017planar,
silva2014performing,backlund2016removing,liu2017huygens,lin2017topology,silva2014performing} not feasible with conventional components. For mid-IR to optical wavelengths, high contrast dielectric metasurfaces are very versatile as they can be designed to control different properties of light on a subwavelength resolution and with large reflection or transmission efficiencies \cite{Fattal2010NatPhoton,kamali2016highly,arbabi2016miniature,
paniagua2017metalens,yang2017multimode,ong2017freestanding,arbabi2015efficient,colburn2017tunable,yang2017topology,
zhou2017efficient,parry2017active,li2017widely,forouzmand2017all,
maguid2017multifunctional}.

Similar to other diffractive devices, metasurfaces that locally control the optical wavefront (e.g. lenses, beam deflectors,  holograms) generally have a fixed response when illuminated from different incident angles, with possible distortions and reduction in efficiency at illumination angles other than the design value \cite{kamali2016,di2011optical,zheng2017wideband}. The main reason for this correlated behavior is the constant locations of the Fresnel zone boundaries (i.e., the generalized grating period) that determine the device function irrespective of the incident angle \cite{Born1999,Fairchild1982}. Moreover, almost in all the demonstrated diffractive and metasurface structures the phase and its local gradient (which is proportional to the local momentum change) have a small dependence on the incident angle~\cite{zheng2017wideband,jang2017complex}, which results in a large optical memory effect range \cite{feng1988correlations}. Here, we introduce angle-multiplexed metasurfaces for simultaneously encoding of different arbitrary phase profiles in different illumination angles of a single sub-wavelength thick metasurface. We introduce a novel angle-dependent platform based on reflective high-contrast dielectric meta-atoms to break the fundamental optical memory effect of metasurfaces and provide independent control over the reflection phase of light at two different incident angles. As a result, any two different functionalities can be embedded in a metasurface that can be separately accessed with different illumination angles. As proof of concept, we experimentally demonstrate angle-multiplexed reflective gratings with different effective grating periods under TE-polarized 0$^{\circ}$ and 30$^{\circ}$ illumination angles (Fig. \ref{fig:1_Concept}b). In addition, we demonstrate an angle-multiplexed hologram which encodes and projects different holographic images under normal and 30$^{\circ}$ illumination angles with TE polarization (Fig. \ref{fig:1_Concept}d).

\begin{figure*}[htp]
\centering
\includegraphics[width=2\columnwidth]{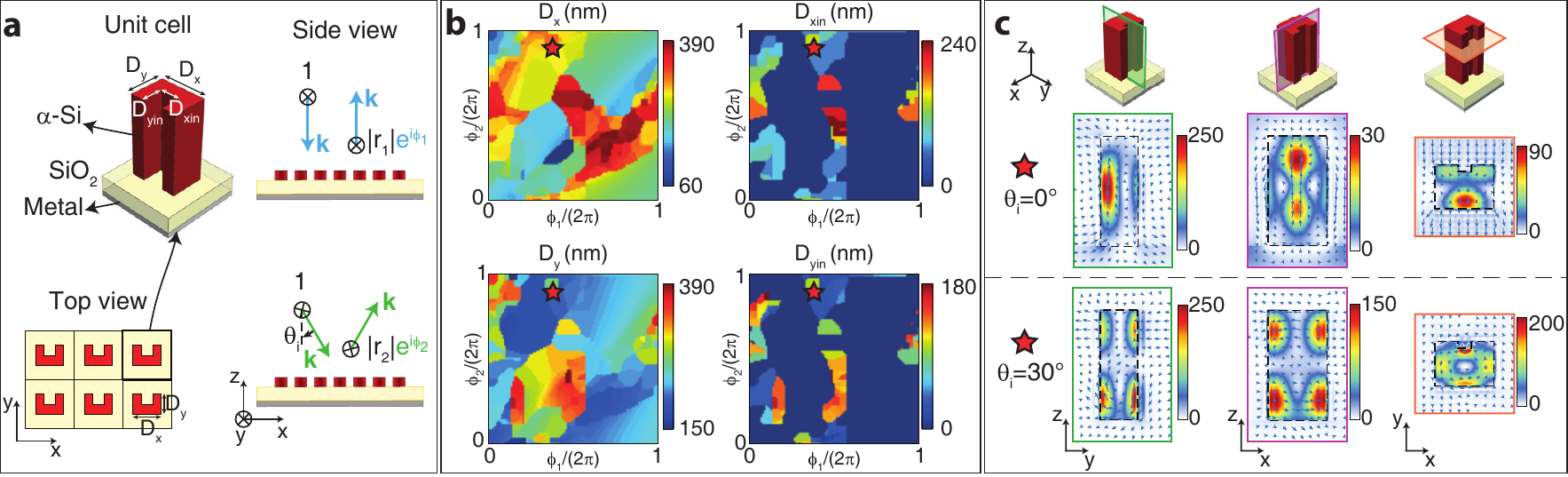}
\caption{\textbf{The meta-atom structure and the design graphs.} \textbf{a}, Schematic drawing of various views of a uniform array of U-shaped
cross-section $\alpha$-Si meta-atoms arranged in a square lattice resting on a thin $\mathrm{SiO_2}$ spacer layer on a reflective surface (i.e., a metallic mirror). The array provides angle-dependent response such that TE-polarized light at 0$^{\circ}$ and 30$^{\circ}$ illumination angles undergo different phase shifts as they reflect from the array. \textbf{b}, Simulated values of the U meta-atom dimensions ($D_x$, $D_y$, $D_{x\mathrm{in}}$, and $D_{y\mathrm{in}}$) for achieving full 2$\pi$ phase shifts for TE-polarized light at 0$^{\circ}$ and 30$^{\circ}$ illumination angles, respectively. One can find the values of the four dimensions of a meta-atom which imposes $\phi_1$ and $\phi_2$ reflection phase shifts onto TE-polarized normal and 30$^{\circ}$ incident angle optical waves from (b). \textbf{c}, Electric energy density inside a single unit cell in a periodic uniform lattice for a typical meta-atom (shown in (b) with a star symbol) at 0$^{\circ}$ and 30$^{\circ}$ illumination angles is plotted in three cross sections. Blue arrows indicate in-plane electric field distributions excited at each illumination angle. Different field distributions at normal and 30$^{\circ}$ incidence is an indication of excitation of different resonant modes under different incident angles. In all parts of the figure, the meta-atoms are 500 nm tall. The silicon dioxide and aluminum layers are 125~nm and 100~nm thick respectively, the lattice constant is 450~nm and all simulations are performed at the wavelength of 915~nm. $\alpha$-Si: amorphous silicon, $\mathrm{SiO_2}$: silicon dioxide.}
\label{fig:2_design}
\end{figure*}

A meta-atom structure capable of providing independent phase control under TE-polarized light illumination with 0$^{\circ}$ and 30$^{\circ}$ incident angles is shown in Fig. \ref{fig:2_design}a. The amorphous silicon ($\alpha$-Si) meta-atoms have a U-shaped cross section (we will call them U meta-atoms from here on) and are located at the vertices of a periodic square lattice on a low refractive index silicon dioxide ($\mathrm{SiO_2}$) and aluminum oxide ($\mathrm{Al_2O_3}$) spacer layers backed by an aluminum reflector. Since the electric field is highly localized in the nano-posts, the low-loss low-index dielectric spacer between the nano-posts and the metallic reflector is necessary to avoid the high losses from metal. In addition, the spacer layer allows for efficient excitation of the resonance modes under both angles of illumination through a constructive interference between the incident and reflected fields inside the nano-posts. Therefore, the nano-posts act as one sided multi-mode resonators \cite{kamali2016,arbabi2017controlling,arbabi2017planar}. For the wavelength of 915~nm, the meta-atoms are 500~nm tall, the $\mathrm{SiO_2}$ layer, the $\mathrm{Al_2O_3}$ layer, and the aluminum reflector are 125~nm, 30~nm, and 100~nm thick, respectively, and the lattice constant is 450~nm. A uniform array of U meta-atoms provides an angle-dependent response such that TE-polarized light waves incident at 0$^{\circ}$ and 30$^{\circ}$ undergo different phase shifts ($\phi_1$ and $\phi_2$, respectively) as they are reflected from the array. A periodic array of U meta-atoms was simulated to find the reflection amplitude and phase at each incident angle (see Appendix A for simulation details). Any combination of $\phi_1$ and $\phi_2$ from 0 to 2$\pi$ can be simultaneously obtained by properly choosing the in-plane dimensions of the meta-atoms (i.e.  $D_x$, $D_y$, $D_{x\mathrm{in}}$, and $D_{y\mathrm{in}}$ as shown in Fig. \ref{fig:2_design}b). Therefore, any two arbitrary and independent phase profiles for TE-polarized 0$^{\circ}$ and 30$^{\circ}$ illumination angles can be designed simultaneously (see Appendix A for design procedure details). The corresponding reflection amplitudes ($|r_1|$ and $|r_2|$) and achieved phase shifts are shown in Supplementary Fig. 1. The independent control of phase at different incident angles is a result of exciting different modes of the U meta-atom under two distinct illumination angles. Figure \ref{fig:2_design}c shows the excited electric energy density for a typical meta-atom in a periodic array at three different cross-sections under 0$^{\circ}$ and 30$^{\circ}$ incident angles (top and bottom receptively). The example meta-atom dimensions and corresponding phases at each illumination angle are shown in Fig. \ref{fig:2_design}b by a star symbol. Modes that are excited under 30$^{\circ}$ illumination angle are different from the excited modes at normal illumination as seen in Fig. \ref{fig:2_design}c. There are two categories of symmetric and antisymmetric resonant modes. In normal incidence only symmetric modes are excited, while in oblique illumination both the symmetric and antisymmetric modes are excited. This is a key factor in realizing this independent control for different angles in a local metasurface platform. As the metasurface is still assumed to be local (i.e., the coupling between adjacent meta-atoms is neglected in the design), any two arbitrary different wavefronts can be simultaneously designed for the two different illumination angles by using the design graphs shown in Fig. \ref{fig:2_design}b. In addition, due to the symmetry of the nano-posts (and also as verified from simulation results) the polarization conversion of the metasurface platform from TE to TM is negligible.

\begin{figure*}[htp]
\centering
\includegraphics[width=2\columnwidth]{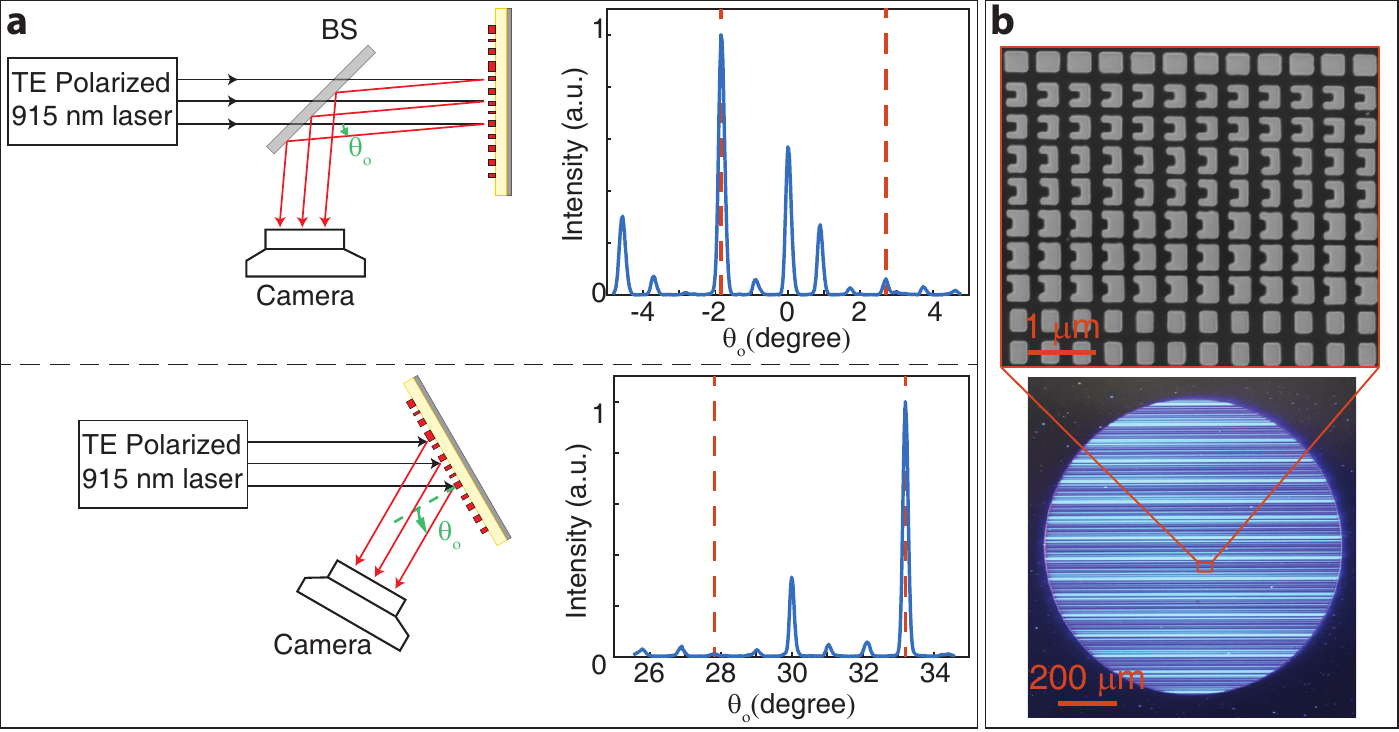}
\caption{\textbf{Angle-multiplexed grating}. \textbf{a}, Simplified schematic of the measurement setup (left), and measured reflectance of the angle-multiplexed grating under normal illumination of TE-polarized light as a function of the observation angle $\theta_\mathrm{0}$ (right). The grating deflects 0$^{\circ}$ and 30$^{\circ}$ TE-polarized incident light to -1.85$^{\circ}$ and +33.2$^{\circ}$ respectively. Orange dashed lines indicate the designed deflection angles (-1.85$^{\circ}$ and +33.2$^{\circ}$ under 0$^{\circ}$ and 30$^{\circ}$ incidence respectively), and the deflection angles corresponding to regular gratings with fixed grating periods (2.7$^{\circ}$ under normal and 27.88$^{\circ}$ under 30$^{\circ}$ illumination angle assuming grating periods of 21$\lambda$ and 31$\lambda$, respectively). See Appendix B and Supplementary Fig. 2
for measurement details. \textbf{b}, Optical image of the angle-multiplexed grating. The inset shows a scanning electron micrograph of the top view of meta-atoms composing the metasurface. See Appendix B for fabrication details. BS: beam splitter.}
\label{fig:3_grating}
\end{figure*}

\begin{figure*}[htp]
\centering
\includegraphics[width=2\columnwidth]{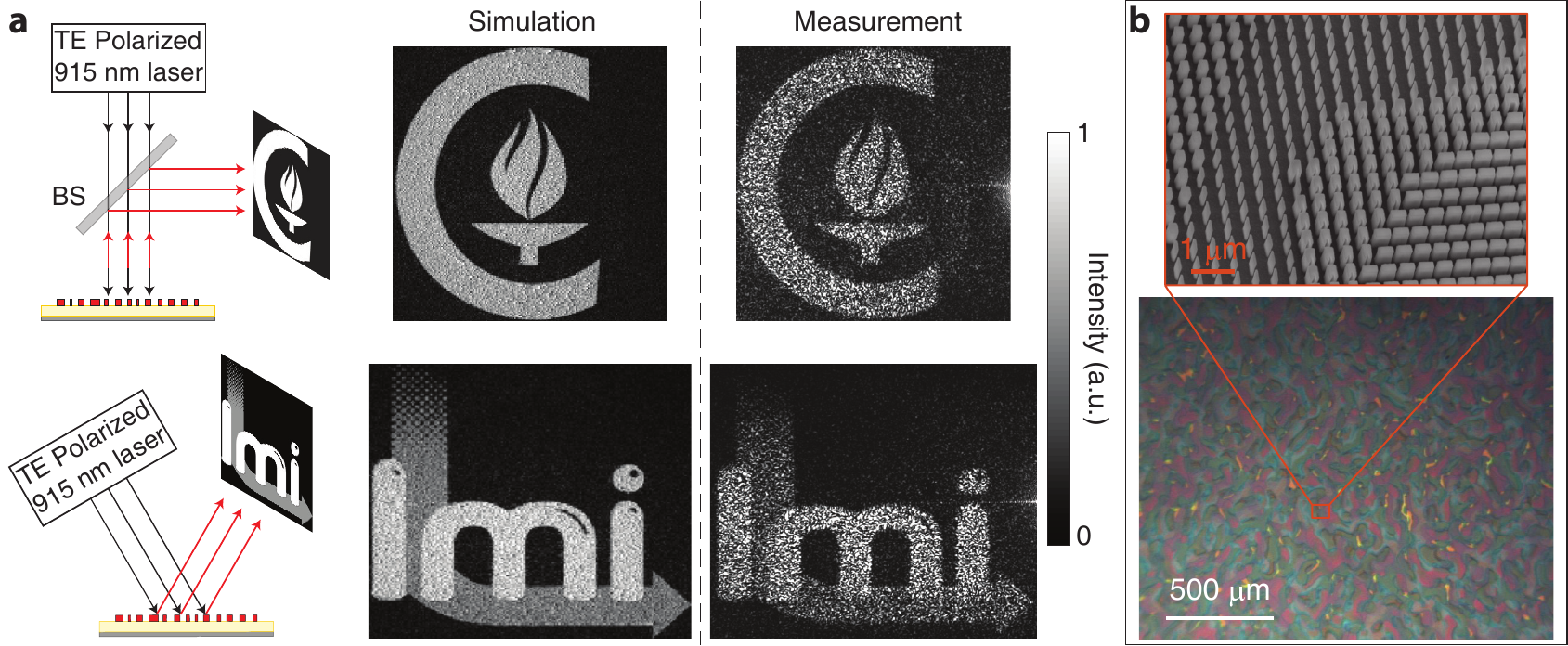}
\caption{\textbf{Angle-multiplexed hologram} \textbf{a}, Simplified drawing of the measurement setups under normal and 30$^{\circ}$ illumination angles (left). The angle-multiplexed hologram is designed to create two different images under different incident angles (Caltech and LMI logos under 0$^{\circ}$ and 30$^{\circ}$, respectively). Simulated and measured reflected images captured under 915-nm TE-polarized light at 0$^{\circ}$ and 30$^{\circ}$ illumination angles are shown on the right. See Appendix B and Supplementary Figs. 3 for measurement details. \textbf{b}, Optical image of a portion of the angle-multiplexed hologram. The inset shows a scanning electron micrograph under oblique view of meta-atoms composing the metasurface. See Appendix B for fabrication details. BS: beam splitter.}
\label{fig:4_hologram}
\end{figure*}

The freedom provided by the proposed platform to simultaneously control the phase of light at two distinct incident angles allows for the implementation of a variety of new compact optical components. To demonstrate the versatility of this platform, we fabricated and characterized two examples of angle-multiplexed metasurfaces. First, an angle-multiplexed grating was designed to operate at 0$^{\circ}$ and 30$^{\circ}$ incident angles with two different effective grating periods. The angle-multiplexed grating has a diameter of 1~mm and deflects 915-nm TE-polarized light incident at 0$^{\circ}$ and 30$^{\circ}$ into -1.85$^{\circ}$ and +33.2$^{\circ}$ respectively. The corresponding effective periods are 31$\lambda$ (blazed for -1 diffraction order) and 21$\lambda$ (blazed for +1 diffraction order) for 0$^{\circ}$ and 30$^{\circ}$ illuminations, respectively ($\lambda=$915~nm is the free space wavelength). The designed devices were fabricated using standard semiconductor fabrication techniques as described in Appendix A. Optical and scanning electron microscope images of the fabricated angle-multiplexed grating are shown in Fig. \ref{fig:3_grating}b. Figure \ref{fig:3_grating}a shows the measured diffracted light intensities versus angle under 0$^{\circ}$ (top) and 30$^{\circ}$ (bottom) TE-polarized illuminations, as well as the simplified measurement setup schematics. The measured reflectance as a function of observation angle shows a dominant peak at the designed angles (i.e. -1.85$^{\circ}$ under normal illumination and +33.2$^{\circ}$ under 30$^{\circ}$ incident angle). Orange dashed lines show deflection angles corresponding to both effective periods, which are 31$\lambda$ (blazed for -1 diffraction order) and 21$\lambda$ (blazed for +1 diffraction order). A regular grating with a 31$\lambda$ period, blazed for -1 diffraction order, would deflect normal incidence into -1.85$^{\circ}$, and 30$^{\circ}$ incident angle into 27.88$^{\circ}$. Similarly, another regular grating with 21$\lambda$ period, blazed for +1 diffraction order, would deflect normal incidence into +2.7$^{\circ}$ and 30$^{\circ}$ incident angle into 33.2$^{\circ}$. The angle-multiplexed grating, on the other hand, deflects 0$^{\circ}$ and 30$^{\circ}$ incident angles into -1.85$^{\circ}$ and +33.2$^{\circ}$ respectively, with no strong deflection peaks at the angle corresponding to the other grating periods (which are, +2.7$^{\circ}$ and 27.88$^{\circ}$). The deflection efficiency of the grating at each incident angle is defined as the power deflected by the grating to the desired order, divided by the power reflected from a plain aluminum reflector (see Appendix B for measurement details and Supplementary Fig. 2 for measurement setups). Deflection efficiencies of 30$\%$ and 41$\%$ were measured under 0$^{\circ}$ and 30$^{\circ}$ incident angles, respectively. For comparison, we simulated the central $\sim$200 $\mu$m-long portion of the grating with a finite difference time domain full-wave electromagnetic solver \citep{Oskooi2010CompPhys} (see Supplementary Note 1 and Supplementary Fig. 3 for simulation results). The simulated deflection efficiencies are 63$\%$ and 54$\%$ for 0$^{\circ}$ and 30$^{\circ}$ operation, respectively. To consider the possible fabrication errors, we also simulated the grating with a random error added to the all in-plane sizes of the meta-atoms. The error is normally distributed with a zero mean, a 4-nm standard deviation, and a forced maximum of 8 nm. The simulated deflection efficiencies with the added errors are 46$\%$ and 39$\%$ under 0$^{\circ}$ and 30$^{\circ}$ incident angles. We attribute the remaining difference between simulated and measured efficiencies to two factors: first, the deposited aluminum reflected layer has a significant surface roughness. This may result in existence and excitation of local surface plasmon resonances that contribute to both increased loss and reflection phase error. Second, to counter the effects of systematic fabrication errors, an array of gratings with different biases added to each size of the meta-atoms is fabricated. In the measurements, one of the devices with good performance under both illumination angles is selected and characterized (i.e., there are other fabricated gratings that demonstrate higher efficiencies for one of the angles). As a result, the characterized device might differ from the one with sizes closest to design values. This may justify the different balances between measured and simulated values for efficiencies under the two illumination angles.

As a second example, an angle-multiplexed hologram which projects two different images under 0$^{\circ}$ and 30$^{\circ}$ illumination angles was designed, fabricated, and characterized. The hologram covers a  2~mm by 2~mm square,  and projects the Caltech and LMI logos when illuminated by TE-polarized light at 915~nm at 0$^{\circ}$ and 30$^{\circ}$ incident angles. Optical and scanning electron microscope images of a portion of the fabricated hologram are shown in Fig. \ref{fig:4_hologram}b. Simulated and measured intensity profiles for two different illumination angles (top and bottom) are shown in Fig. \ref{fig:4_hologram}a, along with simplified schematics of the measurement setups. The Caltech logo is created under normal illumination. By scanning the incident angle from 0$^{\circ}$ to 30$^{\circ}$, the projected image changes from the Caltech logo to the LMI logo. The change in the recorded image with incident angle is shown in Supplementary Movie 1. The good agreement between the simulation and measurement results confirms the independent control of this platform over distinct incident angles. In order to avoid an overlap between the holographic image and the zeroth-order diffraction, the holograms are designed to operate off axis (see Appendix A for details of hologram design). 

The angle-multiplexed metasurface
platform allows for devices that perform completely
independent functions (i.e. grating, lens, hologram, orbital
angular momentum generator, etc.) for different angles of illumination. It is worth noting that the concept and implementation of the angle-multiplexed metasurfaces are fundamentally different from multi-order gratings. While the multi-order gratings can be designed such that the efficiencies of different diffraction orders vary with the incident angle \cite{cheng2017optimization,asadchy2017flat}, the grating momentum corresponding to each order (which is locked to the period of the grating) remains fixed. This difference becomes much clearer when considering the case of holograms. Unlike in the demonstrated platform, it is not possible to encode two completely independent phase profiles corresponding to two completely independent functions in a multi-order holographic optical element (i.e., the generalized case of the multi-order gratings).

In conclusion, we developed optical metasurfaces that break the angular correlation of thin diffractive components, and enable devices where independent phase masks can be embedded in a single thin layer and accessed separately under different illumination angles. Here, the shape of the meta-atom was chosen intuitively and we expect that by utilizing more advanced optimization procedures, the independent control can be extended to more angles and the device performance can be improved significantly. From a technological point of view, this is a novel class of metasurfaces that opens the path towards ultracompact multifunctional flat devices not feasible otherwise. This is complementary to the previously demonstrated independent control over different polarizations \cite{Arbabi2015NatNano,mueller2017metasurface} or wavelengths of the incident light  \cite{arbabi2016high,Aieta2015Science,arbabi2016multiwavelength2,lin2016photonic}, and thus significantly expands the range of applications for nano-engineered metasurfaces.

\section*{Acknowledgement}
This work was supported by the DOE "Light-Material Interactions in Energy Conversion" Energy Frontier Research Center funded by the US Department of Energy, Office of Science, Office of Basic Energy Sciences under Award no. DE-SC0001293. A.A., E.A., and M.F. were supported by Samsung Electronics. A.A. and Y.H were also supported by DARPA. The device nanofabrication was performed at the Kavli Nanoscience Institute at Caltech.

\section*{APPENDIX A: Simulation and Design}

To find the reflection amplitude and phase of a uniform array of meta-atoms, the rigorous coupled wave analysis (RCWA) technique was used \cite{Liu2012CompPhys}. A normal and a 30$^{\circ}$ incident plane wave at 915~nm wavelength were used as the excitation, and the amplitude and phase of the reflected wave were extracted. The subwavelength lattice for both normal and oblique illumination angles results in the excitation of only the zeroth order diffracted light. This justifies the use of only one reflection value at each illumination angle for describing the optical behavior of the meta-atom at each illumination angle. The $\alpha$-Si layer was assumed to be 500~nm thick. The $\mathrm{SiO_2}$ and aluminum layers were assumed to be 125~nm and 100~nm thick, respectively. Refractive indices at 915~nm wavelength were assumed as follows: $\alpha$-Si: 3.558, SiO2: 1.44, Al2O3: 1.7574, and Al: 1.9183-$i$8.3447. The meta-atom in-plane dimensions ($D_x$, $D_y$, $D_{x\mathrm{in}}$, and $D_{y\mathrm{in}}$) are swept such that the minimum feature size remains larger than 50~nm for relieving fabrication constraints.

The optimum meta-atom dimensions for each lattice site at the two incident angles were found by minimizing the total reflection error, which is defined as $\mathrm{\epsilon = |exp(i\phi_1) - r_1|^2 + |exp(i\phi_2) - r_2|^2}$, where $\mathrm{r_1}$ and $\mathrm{r_2}$ are the complex reflection coefficients of the unit-cell at the two incident angles. Therefore, for any desired combination of phases $\phi_1$ and $\phi_2$ in the 0 to 2$\pi$ range at the two incident angles, there is a corresponding meta-atom (i. e., $D_x$, $D_y$, $D_{x\mathrm{in}}$, and $D_{y\mathrm{in}}$ values) that minimizes the reflection error. To limit the rapid jumps in dimensions shown in Fig.\ref{fig:2_design}b, some modification terms  were added to the reflection error in order to ensure that adjacent dimensions are preferred for the adjacent phases. The modification terms were defined as an exponential function of the Euclidean distance between the in-plane dimensions of the meta-atoms for adjacent phase values. 

The holograms of different incident angles were designed individually using the Gerchberg-Saxton (GS) algorithm with $\sim3^\circ$ deflection angles. The simulation results presented in Fig. \ref{fig:4_hologram} were computed by assuming that the coupling among adjacent meta-toms are negligible, such that each meta-atom imposes the exact complex reflection amplitude found from simulations of the periodic structure. The hologram area was assumed to be illuminated uniformly with 0$^{\circ}$ and 30$^{\circ}$ incident angle plane waves, and the projected holographic images were found by taking the Fourier transform of the field after being reflected from the phase mask.

\section*{APPENDIX B: Sample fabrication and Measurement procedure}

A $\sim$100-nm aluminum layer was evaporated on a silicon wafer, followed by a $\sim$30-nm $\mathrm{Al_2O_3}$ layer. A 125-nm-thick $\mathrm{SiO_2}$ and a 500-nm-thick $\alpha$-Si layer were subsequently deposited using the plasma enhanced chemical vapor deposition (PECVD) technique at $200^\circ$C. A Vistec EBPG5200 e-beam lithography system was used to define the pattern in a $\sim$300 nm thick layer of ZEP-520A positive electron-beam resist (spin coated at 5000 rpm for 1 min). The pattern was developed in the resist developer (ZED-N50 from Zeon Chemicals) for 3 minutes. A $\sim$50-nm-thick $\mathrm{Al_2O_3}$ layer was evaporated on the sample, and the pattern was then transferred to the $\mathrm{Al_2O_3}$ layer by a lift off process. The patterned $\mathrm{Al_2O_3}$ hard mask was then used to dry etch the $\alpha$-Si layer in a mixture of $\mathrm{SF_6}$ and $\mathrm{C_4F_8}$ plasma. Finally, the $\mathrm{Al_2O_3}$ mask was removed in a 1:1 solution of ammonium hydroxide and hydrogen peroxide at $80^\circ$C.    

The angle-multiplexed grating was measured using the setup shown schematically in Supplementary Fig. S2. A 915-nm fiber-coupled semiconductor laser was used for illumination and a fiber collimation package (Thorlabs F220APC-780) was used to collimate the incident beam. A polarizer (Thorlabs LPVIS100-MP2) was inserted to confirm the TE polarization state of the incident light. An additional lens with a focal length of 10 cm (Thorlabs AC254-100-B-ML) was placed before the grating at a distance of $\sim$8 cm to partially focus the beam and reduce the beam divergence after being deflected by the grating in order to decrease the measurement error. The light deflected from the device was imaged using a custom built microscope. The microscope consists of a 10X objective lens (Mitutoyo M Plan Apo 10X, NA= 0.28) and a tube lens (Thorlabs LB1945-B-ML) with a focal distance of 20~cm, which images the object plane onto a camera (CoolSNAP K4 from Photometrics). A rotation stage was used to adjust the illumination angle and a 50/50 beamsplitter (Thorlabs NIR Non-Polarizing Cube Beamsplitter) was inserted before the grating for measurements under normal illumination. For efficiency measurements of the grating, an iris was used to select the desired diffraction order and block all other diffraction orders. A power meter (Thorlabs PM100D) with a photodetector (Thorlabs S122C) was used to measure the deflected power off the grating, as well as the reflected power from a plain aluminum reflector (from an area adjacent to the grating). The grating efficiency was calculated by dividing the power deflected to the desired order to the power reflected by the aluminum reflector. Neutral density (ND) filters (Thorlabs ND filters, B coated) were used to adjust the light intensity and decrease the background noise captured by the camera.

The angle-multiplexed hologram was characterized using the setup shown schematically in Supplementary Fig. S3. The setup is similar to the grating measurement setup with some modifications. The 10~cm focal distance lens used to partially focus light to the grating was removed to obtain a relatively uniform illumination of the hologram area. The input beam being larger than the device in addition to fabrication imperfections results in a strong zeroth-order diffraction. The zeroth-order diffraction is cropped in Fig.\ref{fig:4_hologram}a, as it is outside the holographic image of interest due to the off-axis design of the hologram. The custom-built microscope was also altered as follows: the objective lens was used to generate a Fourier transform of the hologram plane in its back focal plane. The tube lens was replaced by a lens with a focal length of 6 cm, which images the back focal plane of the objective into the camera. Two rotation stages were used in order to be able to independently rotate the device and the illumination beam. The camera and the imaging setups were not on the rotation stages.

\clearpage

\newcommand{\beginsupplement}{%
        \setcounter{table}{0}
        \renewcommand{\thetable}{S\arabic{table}}%
        \setcounter{figure}{0}
        \renewcommand{\thefigure}{S\arabic{figure}}%
     }

\onecolumngrid

      \beginsupplement

\section{Supplementary Note 1: Angle-multiplexed grating simulation results}

The central $\sim$200-$\mu$m-long portion of the grating presented in the main text, was simulated for comparison. The simulated grating is 445 lattice constants long in the $x$ direction and 1 lattice constant long in the $y$ direction. Periodic boundary condition was considered in the $y$ direction. The grating was simulated at the wavelength of 915~nm in MEEP \cite{oskooi2010meep} and normal and $30^\circ$ incident y-polarized (TE) plane-waves were used as the excitation. Angular distribution of the reflected power at $0^\circ$ and $30^\circ$ incident angles are shown in supplementary Figs. 3a and 3b, respectively. The far field reflected power was analyzed by taking the Fourier transform of the reflected field above the meta-atoms. The deflection efficiency was calculated by dividing the deflected power to the desired order by the total input power. The simulated deflection efficiency for $0^\circ$ and $30^\circ$ incident angles were 63$\%$ and 54$\%$ respectively. Existence of no other strong diffraction order in supplementary Figs. 3a and 3b, and the high deflection efficiencies achieved demonstrate the independent control of the platform at each incident angle. To consider the possible fabrication errors, the grating with a random error added to all the in-plane sizes of the meta-atoms is also simulated. The error is normally distributed with a zero mean, a 4-nm standard deviation, and a forced maximum of 8 nm. Angular distribution of the reflected power at $0^\circ$ and $30^\circ$ incident angles for the grating with a random error are shown in supplementary Figs. 3c and 3d, respectively. The simulated deflection efficiencies with the added errors are 46$\%$ and 39$\%$ under $0^\circ$ and $30^\circ$ incident angles. Although the deflection efficiency of the grating drops by adding a random random, its general functionality remains the same according to the supplementary Figs. 3c and 3d.

\clearpage
\section{Supplementary Figures}
\begin{figure*}[htp]
\centering
\includegraphics[width=0.9\columnwidth]{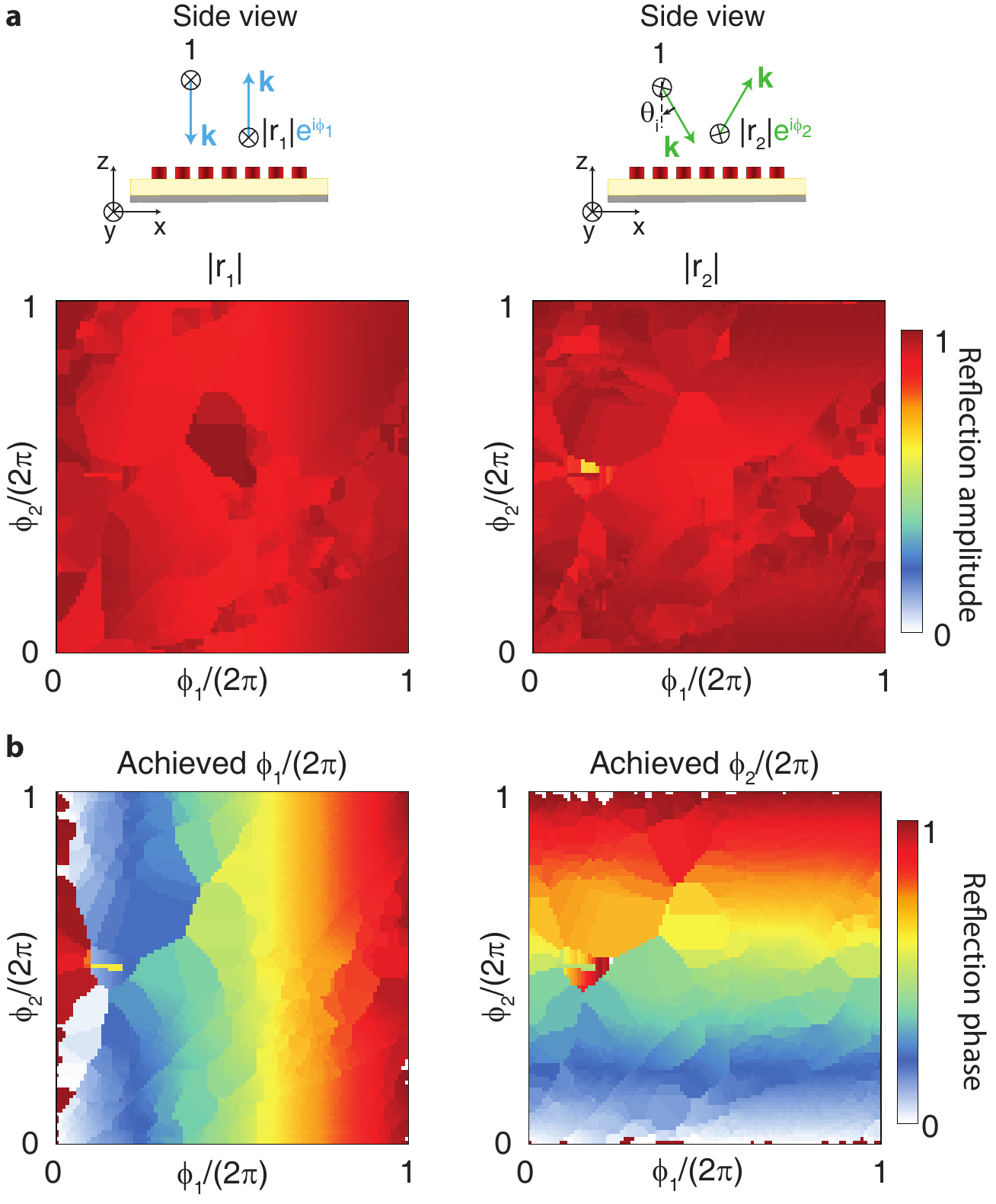}
\caption{\textbf{Simulated achieved reflection amplitudes and phases for the selected meta-atoms.} \textbf{a}, Simulated reflection amplitudes at $0^\circ$ and $30^\circ$ incident angles as a function of required phase shifts for the periodic array of selected meta-atoms that can span the full 2$\pi$ by 2$\pi$ phases for both incident angles. \textbf{b}, Simulated achieved phase shifts of the chosen nano-posts versus the required phase shift values.}
\end{figure*}

\begin{figure*}[htp]
\centering
\includegraphics[width=\columnwidth]{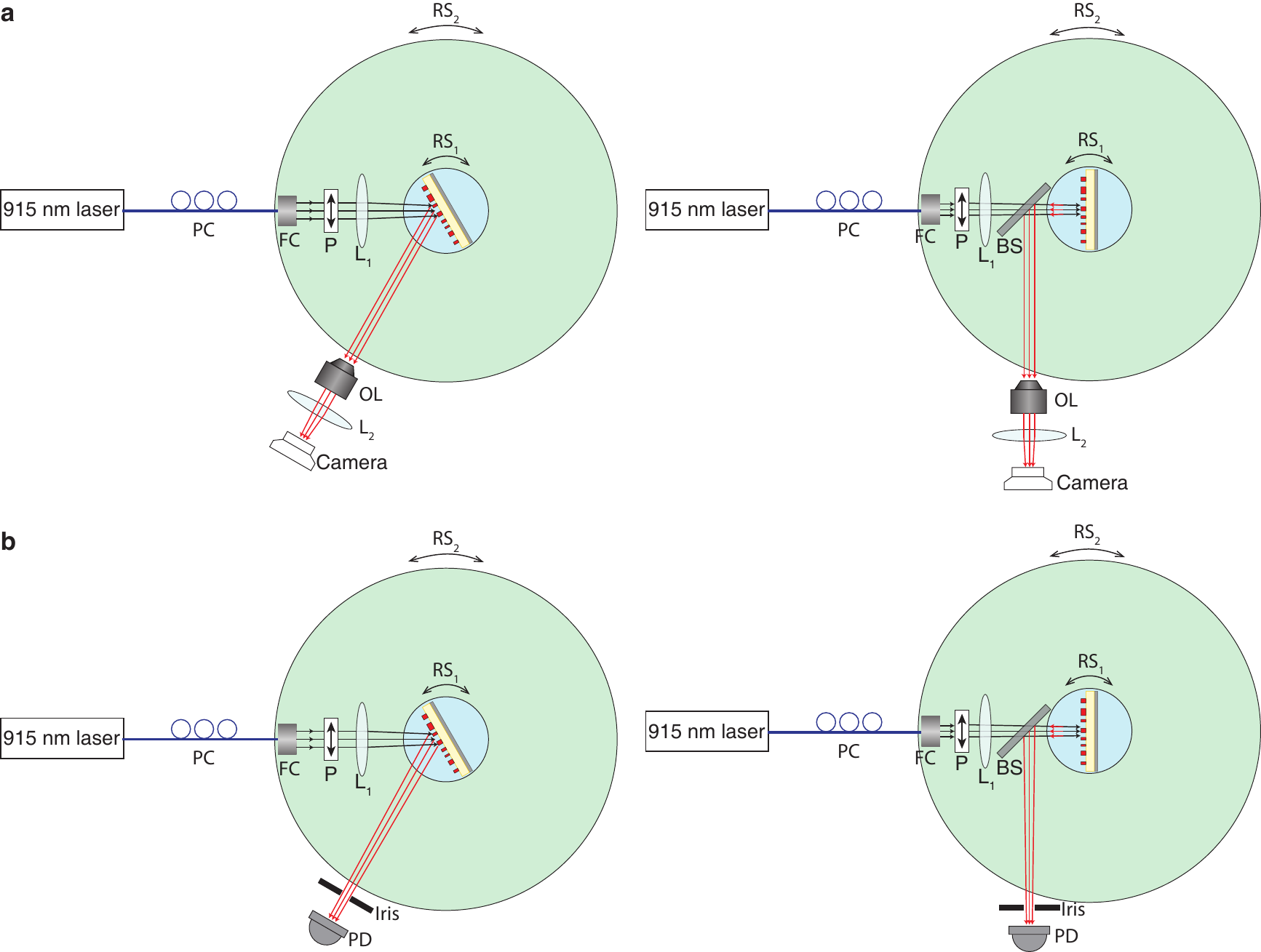}
\caption{\textbf{Measurement setup used to characterize the grating.} \textbf{a}, Schematic drawing of the measurement setup used for characterization of the grating under oblique (left) and normal (right) illumination angles. \textbf{b}, Schematic illustration of the measurement setup used for characterization of deflection efficiency for oblique (left) and normal (right) illuminations. BS: beam splitter, L: lens, PC: polarization controller, FC: fiber collimator, P: polarizer, PD: photodetector. RS: rotation stage. OL: objective lens. The focal lengths of lenses $\mathrm{L_1}$ and $\mathrm{L_2}$ are $f_1=10~\mathrm{cm}$ and $f_2=20~\mathrm{cm}$, respectively.}
\end{figure*}

\clearpage
\begin{figure*}[htp]
\centering
\includegraphics[width=\columnwidth]{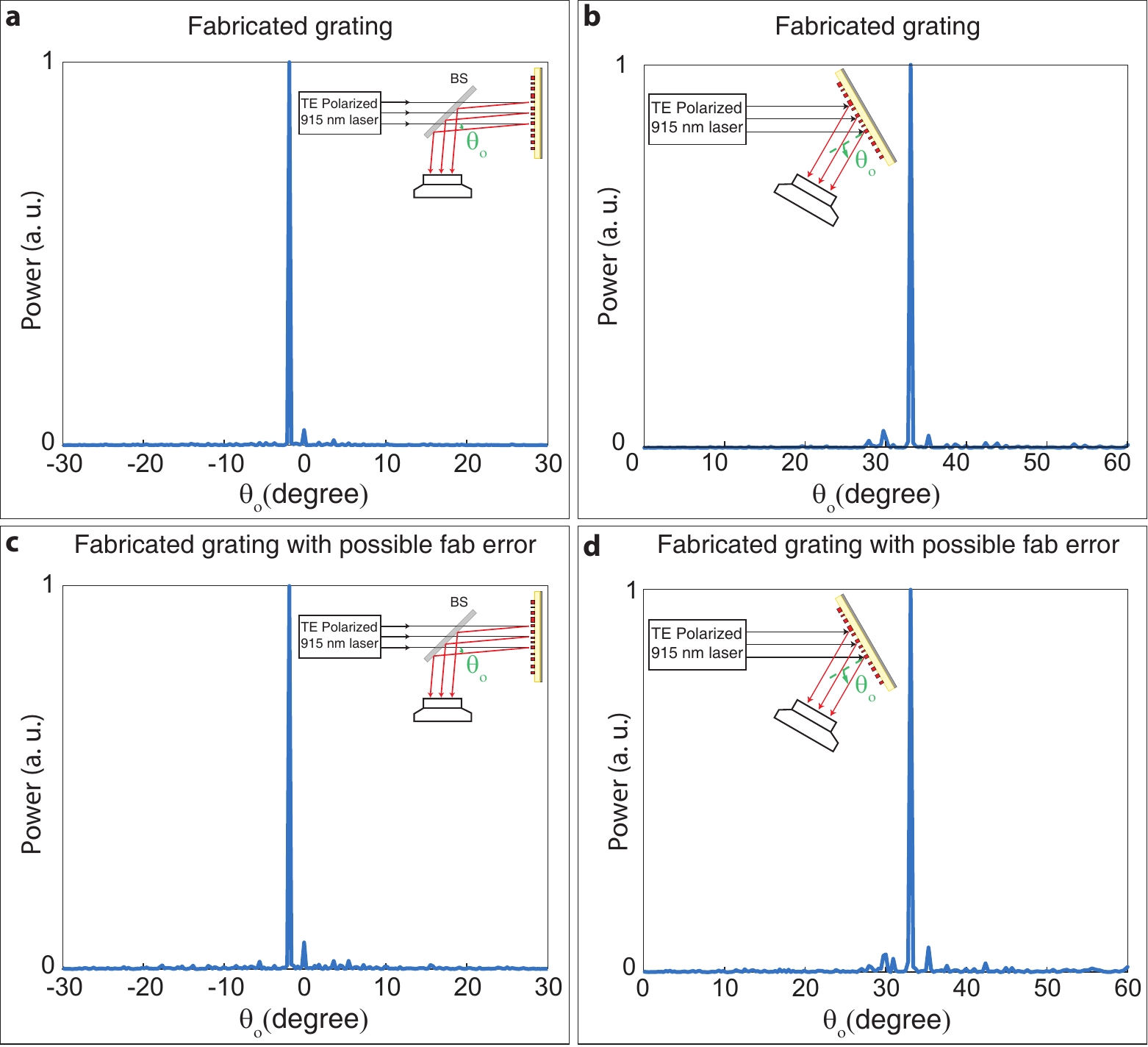}
\caption{\textbf{Simulation results of the angle-multiplexed grating.} \textbf{a} and \textbf{b}, Distribution of reflected power versus observation angle under $0^\circ$ (a) and $30^\circ$ (b) incident angles for a $\sim$200-$\mu$m-long portion of the fabricated grating. \textbf{c} and \textbf{d}, The same graphs as (a) and (b), but with a random error added to the all in-plane sizes of the meta-atoms. The error is normally distributed with a zero mean, a 4-nm standard deviation, and a forced maximum of 8 nm.}
\end{figure*}

\clearpage
\begin{figure*}[htp]
\centering
\includegraphics[width=\columnwidth]{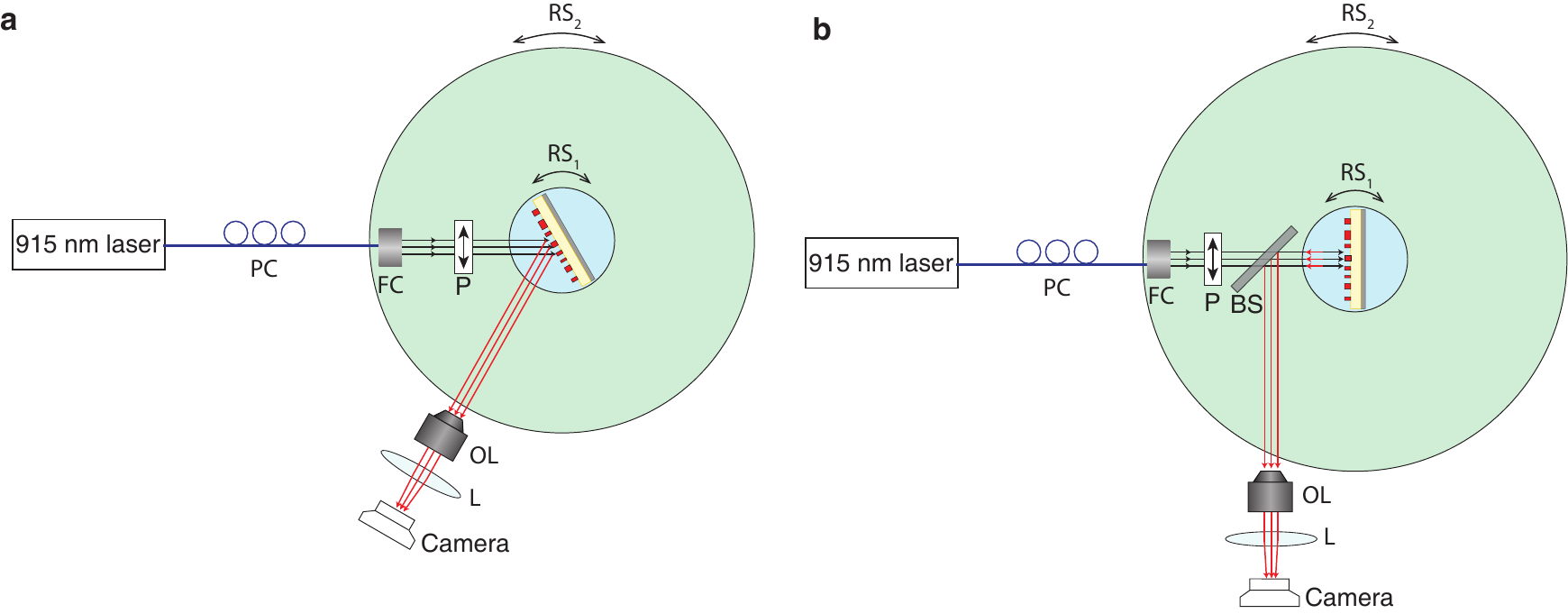}
\caption{\textbf{Measurement setup used for the hologram.} \textbf{a} and \textbf{b}, Schematic drawing of the measurement setup used for characterization of the hologram under oblique (a) and normal (b) illumination angles. BS: beam splitter, L: lens, PC: polarization controller, FC: fiber collimator, P: polarizer, PD: photodetector. RS: rotation stage. OL: objective lens. The focal length of lens L is $f_1=6~\mathrm{cm}$.}
\end{figure*}

\section{Supplementary References}

\end{document}